
\def\etal{{\it et al.}\hskip 1.5pt}
\def\cf{{\it cf.}\hskip 1.5pt}
%
\newbox\hdbox%
\newcount\hdrows%
\newcount\multispancount%
\newcount\ncase%
\newcount\ncols
\newcount\nrows%
\newcount\nspan%
\newcount\ntemp%
\newdimen\hdsize%
\newdimen\newhdsize%
\newdimen\parasize%
\newdimen\thicksize%
\newdimen\thinsize%
\newdimen\tablewidth%
\newif\ifcentertables%
\newif\ifendsize%
\newif\iffirstrow%
\newif\iftableinfo%
\newtoks\dbt%
\newtoks\hdtks%
\newtoks\savetks%
\newtoks\tableLETtokens%
\newtoks\tabletokens%
\newtoks\widthspec%
%
%
\immediate\write15{%
-----> TABLE MACROS LOADED%
}%
%
%
\tableinfotrue%
\catcode`\@=11
\def\tstrut{\vrule height16pt depth6pt width0pt}%
\def\|{|}
\def\tablerule{\noalign{\hrule height\thinsize depth0pt}}%
\thicksize=1.5pt
\thinsize=0.6pt
\def\thickrule{\noalign{\hrule height\thicksize	depth0pt}}%
\def\ctr#1{\hfil\ #1\hfil}%
%
%
%
\tablewidth=-\maxdimen%
\def\tabskipglue{0pt plus 1fil minus 1fil}%
%
%
\centertablestrue%
%
%
%
%
\parasize=4in%
\gdef\ARGS{########}
\gdef\headerARGS{####}
\def\@mpersand{&}
{\catcode`\|=13
\gdef\letbarzero{\let|0}
\gdef\letbartab{\def|{&&}}
}
{\def\ampskip{&\omit\hfil&}
\catcode`\&=13
\let&0
\xdef\letampskip{\def&{\ampskip}}
}
\def\begintable{
   \begingroup%
   \catcode`\|=13\letbartab%
   \catcode`\&=13\letampskip%
   \def\multispan##1{
      \omit \mscount##1%
      \multiply\mscount\tw@\advance\mscount\m@ne%
      \loop\ifnum\mscount>\@ne \sp@n\repeat%
   }
   \def\|{%
      &\omit\widevline&%
   }%
   \ruledtable
}
\long\def\ruledtable#1\endtable{%
%
%
%
   \offinterlineskip
   \tabskip 0pt
   \def\widevline{\vrule width\thicksize}
   \def\endrow{\@mpersand\omit\hfil\crnorm\@mpersand}%
   \def\crthick{\@mpersand\crnorm\thickrule\@mpersand}%
   \def\crnorule{\@mpersand\crnorm\@mpersand}%
   \let\nr=\crnorule
   \def\endtable{\@mpersand\crnorm\thickrule}%
   \let\crnorm=\cr
%
%
   \edef\cr{\@mpersand\crnorm\tablerule\@mpersand}%
   \the\tableLETtokens
%
%
   \tabletokens={&#1}
%
%
   \countROWS\tabletokens\into\nrows%
   \countCOLS\tabletokens\into\ncols%
%
%
   \advance\ncols by -1%
   \divide\ncols by 2%
   \advance\nrows by 1%
%
%
   \iftableinfo	%
      \immediate\write16{[Nrows=\the\nrows, Ncols=\the\ncols]}%
   \fi%
%
%
   \ifcentertables
      \line{
      \hss
   \else %
      \hbox{%
   \fi
      \vbox{%
	 \makePREAMBLE{\the\ncols}
	 \edef\next{\preamble}
	 \let\preamble=\next
	 \makeTABLE{\preamble}{\tabletokens}
      }
      \ifcentertables \hss}\else }\fi
   \endgroup
   \tablewidth=-\maxdimen
}
\def\makeTABLE#1#2{
   {
   \let\ifmath0
   \let\header0
   \let\multispan0
%
%
   \ifdim\tablewidth>-\maxdimen	%
 \widthspec=\expandafter{\expandafter t\expandafter o%
 \the\tablewidth}%
   \else %
      \widthspec={}%
   \fi %
   \xdef\next{
      \halign\the\widthspec{%
      #1
      \noalign{\hrule height\thicksize depth0pt}
      \the#2\endtable
%
      }
   }
   }
   \next
}
\def\makePREAMBLE#1{
   \ncols=#1
   \begingroup
   \let\ARGS=0
   \edef\xtp{\widevline\ARGS\tabskip\tabskipglue%
   &\tstrut\ctr{\ARGS}}
   \advance\ncols by -1
   \loop
      \ifnum\ncols>0 %
      \advance\ncols by	-1%
      \edef\xtp{\xtp&\vrule width\thinsize\ARGS&\ctr{\ARGS}}%
   \repeat
   \xdef\preamble{\xtp&\widevline\ARGS\tabskip0pt%
   \crnorm}
   \endgroup
}
\def\countROWS#1\into#2{
   \let\countREGISTER=#2%
   \countREGISTER=0%
   \expandafter\ROWcount\the#1\endcount%
}%
\def\ROWcount{%
   \afterassignment\subROWcount\let\next= %
}%
\def\subROWcount{%
   \ifx\next\endcount %
      \let\next=\relax%
   \else%
      \ncase=0%
      \ifx\next\cr %
	 \global\advance\countREGISTER by 1%
	 \ncase=0%
      \fi%
      \ifx\next\endrow %
	 \global\advance\countREGISTER by 1%
	 \ncase=0%
      \fi%
      \ifx\next\crthick	%
	 \global\advance\countREGISTER by 1%
	 \ncase=0%
      \fi%
      \ifx\next\crnorule %
	 \global\advance\countREGISTER by 1%
	 \ncase=0%
      \fi%
      \ifx\next\header %
	 \ncase=1%
      \fi%
      \relax%
      \ifcase\ncase %
	 \let\next\ROWcount%
      \or %
	 \let\next\argROWskip%
      \else %
      \fi%
   \fi%
   \next%
}
\def\counthdROWS#1\into#2{%
\dvr{10}%
   \let\countREGISTER=#2%
   \countREGISTER=0%
\dvr{11}%
\dvr{13}%
   \expandafter\hdROWcount\the#1\endcount%
\dvr{12}%
}%
\def\hdROWcount{%
   \afterassignment\subhdROWcount\let\next= %
}%
\def\subhdROWcount{%
   \ifx\next\endcount %
      \let\next=\relax%
   \else%
      \ncase=0%
      \ifx\next\cr %
	 \global\advance\countREGISTER by 1%
	 \ncase=0%
      \fi%
      \ifx\next\endrow %
	 \global\advance\countREGISTER by 1%
	 \ncase=0%
      \fi%
      \ifx\next\crthick	%
	 \global\advance\countREGISTER by 1%
	 \ncase=0%
      \fi%
      \ifx\next\crnorule %
	 \global\advance\countREGISTER by 1%
	 \ncase=0%
      \fi%
      \ifx\next\header %
	 \ncase=1%
      \fi%
\relax%
      \ifcase\ncase %
	 \let\next\hdROWcount%
      \or%
	 \let\next\arghdROWskip%
      \else %
      \fi%
   \fi%
   \next%
}%
{\catcode`\|=13\letbartab
\gdef\countCOLS#1\into#2{%
   \let\countREGISTER=#2%
   \global\countREGISTER=0%
   \global\multispancount=0%
   \global\firstrowtrue
   \expandafter\COLcount\the#1\endcount%
   \global\advance\countREGISTER by 3%
   \global\advance\countREGISTER by -\multispancount
}%
\gdef\COLcount{%
   \afterassignment\subCOLcount\let\next= %
}%
{\catcode`\&=13%
\gdef\subCOLcount{%
   \ifx\next\endcount %
      \let\next=\relax%
   \else%
      \ncase=0%
      \iffirstrow
	 \ifx\next& %
	    \global\advance\countREGISTER by 2%
	    \ncase=0%
	 \fi%
	 \ifx\next\span	%
	    \global\advance\countREGISTER by 1%
	    \ncase=0%
	 \fi%
	 \ifx\next| %
	    \global\advance\countREGISTER by 2%
	    \ncase=0%
	 \fi
	 \ifx\next\|
	    \global\advance\countREGISTER by 2%
	    \ncase=0%
	 \fi
	 \ifx\next\multispan
	    \ncase=1%
	    \global\advance\multispancount by 1%
	 \fi
	 \ifx\next\header
	    \ncase=2%
	 \fi
	 \ifx\next\cr	    \global\firstrowfalse \fi
	 \ifx\next\endrow   \global\firstrowfalse \fi
	 \ifx\next\crthick  \global\firstrowfalse \fi
	 \ifx\next\crnorule \global\firstrowfalse \fi
      \fi
\relax
      \ifcase\ncase %
	 \let\next\COLcount%
      \or %
	 \let\next\spancount%
      \or %
	 \let\next\argCOLskip%
      \else %
      \fi %
   \fi%
   \next%
}%
\gdef\argROWskip#1{%
   \let\next\ROWcount \next%
}
\gdef\arghdROWskip#1{%
   \let\next\ROWcount \next%
}
\gdef\argCOLskip#1{%
   \let\next\COLcount \next%
}
}
}
\def\spancount#1{
   \nspan=#1\multiply\nspan by 2\advance\nspan by -1%
   \global\advance \countREGISTER by \nspan
   \let\next\COLcount \next}%
\def\dvr#1{\relax}%
\def\header#1{%
\dvr{1}{\let\cr=\@mpersand%
\hdtks={#1}%
\counthdROWS\hdtks\into\hdrows%
\advance\hdrows	by 1%
\ifnum\hdrows=0	\hdrows=1 \fi%
\dvr{5}\makehdPREAMBLE{\the\hdrows}%
\dvr{6}\getHDdimen{#1}%
{\parindent=0pt\hsize=\hdsize{\let\ifmath0%
\xdef\next{\valign{\headerpreamble #1\crnorm}}}\dvr{7}\next\dvr{8}%
}%
}\dvr{2}}
\def\makehdPREAMBLE#1{
\dvr{3}%
\hdrows=#1
{
\let\headerARGS=0%
\let\cr=\crnorm%
\edef\xtp{\vfil\hfil\hbox{\headerARGS}\hfil\vfil}%
\advance\hdrows	by -1
\loop
\ifnum\hdrows>0%
\advance\hdrows	by -1%
\edef\xtp{\xtp&\vfil\hfil\hbox{\headerARGS}\hfil\vfil}%
\repeat%
\xdef\headerpreamble{\xtp\crcr}%
}
\dvr{4}}
\def\getHDdimen#1{%
\hdsize=0pt%
\getsize#1\cr\end\cr%
}
\def\getsize#1\cr{%
\endsizefalse\savetks={#1}%
\expandafter\lookend\the\savetks\cr%
\relax \ifendsize \let\next\relax \else%
\setbox\hdbox=\hbox{#1}\newhdsize=1.0\wd\hdbox%
\ifdim\newhdsize>\hdsize \hdsize=\newhdsize \fi%
\let\next\getsize \fi%
\next%
}%
\def\lookend{\afterassignment\sublookend\let\looknext= }%
\def\sublookend{\relax%
\ifx\looknext\cr %
\let\looknext\relax \else %
   \relax
   \ifx\looknext\end \global\endsizetrue \fi%
   \let\looknext=\lookend%
    \fi	\looknext%
}%
%
%
\def\tablelet#1{%
   \tableLETtokens=\expandafter{\the\tableLETtokens #1}%
}%
\catcode`\@=12
\def\refset{\parindent=0pt\hangafter=1\hangindent=1em}
\magnification=1200
\hsize=6.00truein
\hoffset=1.20truecm
\newcount\eqtno
\eqtno = 1
\parskip 3pt plus 1pt minus .5pt
\baselineskip 20pt plus .1pt
\centerline{ \  }
\vskip 0.35in
\centerline{\bf INTERNAL VELOCITY AND MASS DISTRIBUTIONS IN}
\centerline{\bf CLUSTERS OF GALAXIES FOR A VARIETY OF COSMOGONIC MODELS}
\vskip 1.0in
\centerline{Renyue Cen}
\centerline{\it Princeton University Observatory}
\centerline{\it Princeton, NJ 08544 USA}
\vskip 2.0in
\centerline{Submitted to {\it The Astrophysical Journal} on Jan 7, 1994}
\vskip 0.3in
\centerline{March 14, 1994}
\vskip 0.7in
\vfill\eject

\centerline{\bf ABSTRACT}
The mass and velocity
distributions in the outskirts ($0.5-3.0h^{-1}$Mpc)
of clusters of galaxies are examined for
a suite of cosmogonic models
utilizing large-scale Particle-Mesh (PM) simulations
($500^3$ cells, $250^3$ particles and box size of $100h^{-1}$Mpc,
giving a nominal resolution of $0.2h^{-1}$Mpc with
the true resolution about $0.5h^{-1}$Mpc).

Through a series of model computations, designed to
isolate the different effects,
we find that both $\Omega_0$ and $P_k$ ($\lambda\le 16h^{-1}$Mpc)
are important to the mass distributions in clusters of galaxies.
There is a correlation between
power, $P_k$,
and density profiles of massive mass clusters;
more power tends to point to the direction
of a correlation between $\alpha$ and $M(r<1.5h^{-1}$Mpc)
[see equation (1) for definitions],
i.e., massive clusters being relatively extended and
small mass clusters being relatively concentrated.
A lower $\Omega_0$ universe tends to produce
relatively concentrated massive clusters and relatively
extended small mass clusters
compared
to their counterparts in a higher $\Omega_0$ model with the same power.
Models with little (initial) small scale power, such as the HDM model,
produce more extended mass distributions
than the isothermal distribution
for most of the clusters.
But the CDM models
show mass distributions of most
of the clusters more concentrated than the isothermal distribution.
X-ray and gravitational lensing
observations
are begining providing useful information
on the mass distribution in and around clusters;
some interesting constraints on
$\Omega_0$ and/or
the (initial) power of the density fluctuations on scales
$\lambda \le 16h^{-1}$Mpc (where linear extrapolation is invalid)
can be obtained when larger observational data sets, such
as the planned Sloan Digital Sky Survey, become available.

With regard to
the velocity distribution, we find two interesting points.
First, in $0.5<r<3.0h^{-1}$Mpc region,
velocity dispersions of four
components, (1d, radial, tangential, line-of-sight),
show decreasing distributions as a function
of cluster-centric distance in the three CDM models;
but the HDM model shows just the opposite: weakly
increasing velocity dispersions outwards.
The CDM models can reasonably fit the observed galaxy velocity
dispersions in the Coma cluster of galaxies but
the HDM model provides a poor fit.
Second, while the velocity dispersions among the three Cartesian
directions are isotropic,
a large scatter (40\%) exists in all models.
We find that for the scales $0.5<r<3.0h^{-1}$Mpc,
the tangential velocity dispersion
is always larger than
the radial component
by a factor of 1.2-1.6 in the CDM models
and 1.3-2.0 in the HDM model.
In all models the ratio of radial to tangential velocity
dispersions is a decreasing function from $0.5h^{-1}$Mpc
to $3.0h^{-1}$Mpc for massive clusters
(smaller clusters tend to show a minimum for that ratio
around $1.5-2.0h^{-1}$Mpc in the CDM models).

We also examine, in detail, the infall problem.
Lower $\Omega_0$ models are found to have
larger turnaround radius for a fixed-mass clump
than high $\Omega_0$ models;
this conclusion is insensitive to $P_k$.
But we find that the following relation
(between the turnaround radius, $R_{ta}$, and
the mass within $R_{ta}$, $M_{ta}$),
$\log_{10}R_{ta} = a + b \log_{10} M_{ta}$
($a=-5.2\pm 0.2$, $b=0.40\pm 0.02$, $R_{ta}$ and $M_{ta}$ are
in $h^{-1}$Mpc and $h^{-1}\hbox{M}_\odot$, respectively),
holds for all the models (the uncertainties in $a$ and $b$
indicate the variations among models).
In addition,
the relation between the overdensity inside the turnaround
radius, $\delta_{ta}$, and $M_{ta}$
is fitted by
$\log_{10}\delta_{ta} = c + d \log_{10} M_{ta}$
(\cf Table 1 for values of $c$ and $d$).
We show that the isolated spherical collapse model
in an Einstein-de Sitter universe,
having $\delta_{ta}=9\pi^2/16=5.55$,
gives a fair fit to results ($\sim 4-10$)
of the nonlinear, non-spherical simulations
performed here.
Lower $\Omega_0$ models have considerably higher
$\delta_{ta}$, $\sim 10-30$.

Finally, we find that
the isothermal approximation (\cf equation 10)
tends to underestimate
the true masses within the Abell radius by
10-30\% with a scatter of $\sim 50\%$ around the
estimated mean (in the three hierachical models).

\noindent
Subject headings:  Cosmology: large-scale structure of Universe
-- cosmology: observations
-- cosmology: theory
-- dark matter
-- galaxies: clustering
\vfill\eject

\centerline{1. INTRODUCTION}

Since the dynamical time of clusters of galaxies
is not much shorter than the Hubble time,
it is expected that they contain useful
information with regard to the early state of the universe.
There are numerous studies from galactic scale to very large, supercluster
scale in a variety of cosmological models
and we will not even pretend to attempt to list them (in vain).
For a recent review of confrontations of a panoply of cosmic
theories with observations see Peebles \& Silk (1988),
for a post-COBE review of the Cold Dark Matter (CDM) model
see Ostriker (1993),
and for an extensive summary report on galaxy formation
and large-scale structure see Silk \& Wyse (1993).
However, our current understanding of the evolution
and properties of clusters of galaxies and
of their relationships with details of cosmological models
is still in its infancy.
One is therefore encouraged to explore any new dimensions.

The central regions of clusters of galaxies
are more relaxed than their outskirts and therefore
are less sensitive to cosmological details.
For example, the mass (within the Abell radius)
function of clusters of galaxies
is found to be dependent on the mean cosmological density,
$\Omega_0$, and the normalization on the relevant scale (e.g., $\sigma_8$),
but not sensitively
on the shape of the power spectrum within physical plausibility
(Bahcall \& Cen 1992; White, Efstathiou, \& Frenk 1993).
The outskirts of the clusters of galaxies ($1.5-3.0h^{-1}$Mpc)
are likely to be more sensitive to
the details of a cosmological model than the central, core regions,
since they have not undergone or have
progressed relatively less
toward virialization and an
equilibrium state.
One would therefore anticipate that the density and velocity fields
in these regions ought to be dependent on such cosmological parameters
as $\Omega_0$, $\sigma_8$, $P_k$.
No sufficiently detailed study on this subject has been done
on the relevant scales.
Motivated by this, this paper is written.
Specifically, we will focus on the velocity and density
fields surrounding the clusters of galaxies on the scales
$0.5-3.0h^{-1}$Mpc.

Traditionally,
the velocity dispersions in clusters of galaxies
are related to their masses.
This interpretation is frequently utilized,
and has become one of the two conventional ways to
determine (based on dynamical grounds) masses of clusters
using observed velocity information
(Peebles 1970;
Rood \etal 1972;
White 1976;
Kent \& Gunn 1982;
Merritt 1986;
The \& White 1986;
Peebles 1993;
Bahcall \& Cen 1993).
The alternative method is to use cluster
X-ray temperature information
(Sarazin 1986;
Cowie, Henriksen, \& Mushotzky 1987;
The \& White 1988;
Hughes 1989;
Henry \& Arnaud 1991;
Bahcall \& Cen 1993), which
is not the subject of this paper.
The \& White (1986) and Merritt (1986)
found
that the inferred mass of the Coma cluster of galaixes
within a radius $1h^{-1}$Mpc is
always very close to $6\times 10^{14} h_{50}^{-1} M_\odot$,
independent of details
with regard to the possible variations of the velocity and mass distributions.
However,
the inferred mass within a radius of $2.7h^{-1}$Mpc
is highly uncertain, ranging from
$6\times 10^{14} h_{50}^{-1} M_\odot$ to
$5\times 10^{15} h_{50}^{-1} M_\odot$.
This large uncertainty is
due to a large range of physically plausible
configurations of the mass
and galaxy velocity, which
are consistent with the observed
line-of-sight velocity dispersions of the galaxies in the Coma cluster.

In this paper we explore the cluster mass distribution
directly using N-body simulations of a variety of cosmogonic
models,
and the velocity distribution under the assumption
that the velocity field of galaxies
in clusters follows that of the underlying mass, i.e.,
there is no velocity ``bias".
But note that the issue of the velocity ``bias"
is still a controversial one.
Different results were found in the work by
Carlberg, Couchman \& Thomas (1990), Carlberg \& Dubinski (1991),
Cen \& Ostriker (1992), Katz, Hernquist, \& Weinberg (1992)
and Evrard, Summers, \& Davis (1994).
But the uncertainty involved is being narrowed down, and at
present the velocity bias value, $b_v\equiv v_{gal}/v_{mass}$,
among different studies
can be described by $b_v=0.85\pm 0.15$.
Hence, although
the previous assumption (there is no velocity bias)
is not necessarily valid (but assumed for the sake of convenience
of comparison)
and a definite conclusion awaits still higher resolution, larger scale,
detailed, hydrodynamic computations with galaxy formation
[for current, state-of-the-art work on this subject see
Cen \& Ostriker (1992, 1993a,b),
Katz, Hernquist, \& Weinberg (1992),
and Evrard, Summers, \& Davis (1994)],
the effect is relatively small (at most 30\%)
even under the present uncertain situation.

In addition, we have to assume that the mass clumps
in a simulation correspond to clusters of galaxies in the real
universe.
While one can not rule out the possibility
that there exist massive, ``dark" clusters in the real universe,
which luminous galaxies happen to like to stay away from
due to whatever physical processes,
it seems unlikely that galaxies can resist enjoying
the safe (deep) potential wells created by such massive clumps.
One way to do this is to assume
that these dark clusters are just formed and the
luminous galaxies (which formed outside of them) have not
had time to migrate in.
But then one can not explain why
such clusters are not X-ray luminous since
the gas in the clusters should have been shock-heated
during the phase of collapse to form the cluster.
Based on these arguments, it seems inprobable
that there exist any ``dark" clusters in the real universe
unless there is some large-scale process which segregates
dark matter from baryons on scales larger than clusters
at the early times.
So we feel that it is a good assumption that mass traces
clusters (either optical or X-ray), i.e., mass clumps
in simulations correspond to clusters of galaxies with similar
masses.
However, we do not resolve dark halos with sizes smaller
than $0.5h^{-1}$ under present simulations, and we do even worse to
tag galaxies in the simulations.
So let us stress once more that,
it is an assumption not necessary a valid statement that
galaxies spatially follow mass in the clusters,
upon which our analyses are based.

We examine
four different cosmogonic models:
1) standard COBE-normalized CDM model with
$(h,\Omega_0,\sigma_8)=(0.5,1.0,1.05)$;
2) standard HDM model with
$(h,\Omega_0,\sigma_8)=(0.5,1.0,1.05)$;
3) an open CDM model with
$(h,\Omega_0,\sigma_8)=(0.5,0.2,1.05)$;
4) an open CDM model with
$(h,\Omega_0,\sigma_8)=(0.5,0.2,1.05)$
but artificially
adopting the $\Omega_0=1$ CDM initial power spectrum.
The first model is a realistic, popular, COBE-normalized cosmogonic model.
The remaining three models are not COBE-normalized
(model 4 is not even physically realistic) but chosen
to examine the parameter space ($\Omega_0, P_k$) with the
desire to isolate the different effects.
The rest of the paper is organized in the following manner.
Section 2 describes the numerical techniques;
\S 3 presents the results;
and \S 4 assembles our conclusions.

\bigskip
\medskip
\centerline{2. METHOD}
\centerline{\it 2.1 Model Simulations}
A standard Particle-Mesh code (PM, \cf Hockney \& Eastwood 1981;
Efstathiou \etal 1985)
with a staggered-mesh scheme (see, e.g., Park 1990; Cen 1992)
is used
to simulate the evolution of the universal matter.
We use $250^3=10^{7.2}$ particles on a $500^3$ mesh with
a periodic, comoving simulation box of size $100h^{-1}$Mpc,
giving a nominal spatial
resolution of $0.2h^{-1}$Mpc.
The gravitational force is calculated by
a FFT technique.
The cloud-in-cell scheme
is used
to assign the gridded gravitational force
to the disordered positions of particles
as well as to calculate the gridded density
from the disordered positions of particles.
We denote a Hubble constant $H_0=100h$km/s/Mpc throughout.

Four models (listed in Table 1) are computed.
Row 3 in Table 1 indicates
the present mean density of the model universe in terms
of the closure density;
row 4 indicates the kind of power spectrum transfer function used;
row 5 is the Hubble constant;
row 6 is the power index of the spectrum on the large-scale end;
row 7 is the mass fluctuation on a top-hat
sphere with a radius $8h^{-1}$Mpc at present
by normalizing the linear power spectrum;
row 8 is simulation box size;
row 9 is simulation cell size;
row 10 indicates the mass of each particle in the simulation.
The last four rows will be described in due course.
We adopt the transfer functions of
Bardeen \etal (1986)
for the CDM and HDM models.
The initial density field for each simulation
is generated assuming Gaussian fluctuations.
The initial velocity field is given by the
Zel'dovich approximation.

\medskip
\centerline{\it 2.2 Cluster Identification}
First, we select out clusters in a simulation
using an adaptive friends-of-friends linking algorithm.
The local linking length $b_{ij}$
between the $i$-th and $j$-th particles is determined by
$
b_{ij} = {\rm Min}[{L_{\rm box}/ N^{1/3}},\beta ({1 \over 2})^{1/3}
({1/n_i(a_s)} +
{1/ n_j(a_s)})^{1/3}],
$
where $L_{\rm box}$ is the box size, $N$ is the total number of
particles in the box,
$n_i(a_s)$ is the local number density at
the $i$-th particle's position smoothed over a Gaussian window
with radius $a_s$.
We use $a_s=10h^{-1}$Mpc and $\beta=0.25$, which are found to
be adequate for our purpose in the sense that it produces neither
too big structures (this will effectively reduce the number of
centers of clusters if, for example,
there are actually two big, adjacent structures which are artificially
lumped together)
nor too small structures (this might lead to treatment of
small systems as rich clusters)
compared to normal clusters of galaxies
(see Bahcall \& Cen 1992).
Experiments with other linking schemes indicate that
the selected list of clusters
does not sensitively depend on
the details of the selection scheme
as long as there is a cutoff radius ($1.5h^{-1}$Mpc in this case)
and one counts all the particles within that radius.
This yields a final list of clusters each
with its center position, where the center of each cluster
is the center of mass of each linked group.
Such defined center of a cluster,
in practice,
always corresponds to the maximum of the central density
peak of a cluster (except in rare cases if there are
multiple dense structures in the central region, i.e.,
substantial substructures exist in the central region).

Second, having defined the centers of the clusters
we return to the simulation box and count
all the particles around each center to a certain radius
(we only consider various properties within
a radius of $3h^{-1}$Mpc in this study).
Typically, each cluster contains hundreds to thousands of
particles. Row 11 in Table 1 indicates the
number of particles contained in a cluster of
mass $1.0\times 10^{14}h^{-1}M_\odot$.
Also relevant is the number density of clusters
in terms of mass (i.e., cluster mass function).
As an illustration, we listed, as row 12 in Table 1,
the cumulative number density of clusters with masses
greater than $1.8\times 10^{14}h^{-1}M_\odot$.
We see that the standard, COBE-normalized, $\Omega_0=1$ CDM model
overproduces such observed clusters by a factor in excess of ten.
In contrast, the two lower $\Omega_0$ CDM models agree well
with observations.
This subject concerning cluster mass function
is very interesting by itself but we will not address
this issue since it has been discussed in depth
in Bahcall \& Cen (1992).

Finally, we examine the properties of internal mass distribution and
velocity distribution of each cluster
as a function of radius
[bin size (thichness of each shell) of
$0.2h^{-1}$Mpc is used, which
is appropriate given our simulation resolution].
Other correlations among these and derived quantities are then also
investigated in detail.
In all cases,
the mass of a certain region
is defined to be the number of particles
contained in that region
multiplied by the mass of each particle.
Note that since each shell ($0.2h^{-1}$Mpc) contains typically
few hundred or more particles, discreteness effect is small.
For example, for an Abell cluster of mass $10^{14}h^{-1}M_\odot$,
each shell will have $(740,3703)$ particles
(assuming the singular isothermal
distribution for the simplicity of illustration)
for $\Omega_0=(1.0,0.2)$ cases,
which translate to Poissonian fluctuations of
(3.7\%,1.6\%), respectively.
Here we focus on the distributions as a function
of radius (averaged over shells)
but do not address the very important
subject of substructures in and around clusters of galaxies.
A subsequent paper will be devoted to this subject
focusing on the dependence of substructures in and around clusters
on $P_k$ and $\Omega_0$.
\vfill\eject

\bigskip
\centerline{3. RESULTS}
\bigskip
The main results are organized into three sections with
regard to the mass distribution, velocity distribution,
and the isothermal model.

\medskip
\centerline{\it 3.1 Cluster Internal Mass Distribution}
\medskip
Figure (1) shows the density distribution (solid curves)
as a function of cluster-centric distance
for a few typical clusters in the four models
[panel (a) for model 1,
panel (b) for model 2,
panel (c) for model 3 and
panel (d) for model 4, see Table 1;
this order will be maintained in the following figures].
Also shown as dashed lines are the slope for an
isothermal sphere.
Note that our nominal resolution (cell size) is
$0.2h^{-1}$Mpc and the true resolution is about 2.5 cells.
The density in the central region of each cluster, $r<0.5h^{-1}$Mpc,
is probably underestimated, due to our limited resolution.
But this defect should not significantly affect the mass distribution
on larger scales.
We see that there is a wide range of density profiles within
each model as well as among the different models.
The three CDM models show a relatively
close range of the asymptotic density slopes on scales $1-3h^{-1}$Mpc,
the HDM model appears to have significantly shallower slopes.
The CDM models have steeper slope than an isothermal
case while the HDM model shows the opposite.

To address the mass distribution in a more quantitative fashion,
we derive the asymptotic slope of the mass distribution around
each cluster by fitting the simulated data points
in the range $1.0 < r < 3.0h^{-1}$Mpc
with a power law
form as
$$\eqalignno{M(<r)&= Ar^\alpha
\quad. \quad &(\the\eqtno )\cr}$$
\advance\eqtno by 1
The relationship between $\alpha$ and $M(r<1.5h^{-1}$Mpc$)$
is shown as open circles in Figure (2) for the four models.
The solid line in each panel
is the linear least-square fit for the open circles
weighted by the inverse of the uncertainty of each power law fit
($\Delta\alpha$).
We see that, in the $\Omega_0=1$ CDM model [panel (a)],
there is a weak trend of more massive clusters being more extended.
In contrast, the HDM model shows a strong anti-correlation.
Also note that there is a concentration around $\alpha=2$ in the HDM
model while it is near $\alpha=0.3$ in the $\Omega_0=1$ CDM model.
The primary reason for these differences ultimately traces to
the fundamental differences of these two scenarios:
the bottom-up picture in the CDM model
and the top-down picture in the HDM model.
The hierachical clustering process (or equivalently,
the process of continuous merging) in the $\Omega_0=1$ CDM model
tends to gradually make massive clusters
more extended due to merging and infalling of satellite objects.
On the contrary, the HDM model
produces highly concentrated mass distribution
in the regions where pancakes intersect;
smaller clusters, which are the products
of fragmentation of big clumps, filaments and sheets,
tend to be more extended and lacking cores.
The two open CDM models [panels (c,d)] appear to be
intermediate between the above two models.
Comparing panel (a) and panel (d) (note that the only difference
between these two models is $\Omega_0$, 1.0 vs. 0.2),
we see that a lower $\Omega_0$
has an effect of producing relatively less extended
massive clusters.
The obvious explanation for this
is that there is significantly less merging in a lower
$\Omega_0$ universe than in an $\Omega_0=1$ universe.
Comparing panel (c) and panel (d)
[note that the only difference
between these two models is the slope of the power spectrum
on the relevant scales;
model (3) (panel c) has a steeper slope of the power spectrum
than model (4) (panel d)],
we find, as a verification of our preceding explanation
for the difference found between $\Omega_0=1$, CDM and HDM models,
that less power (a steeper slope in model 3 than in model 4)
on the relevant
scales indeed points to the direction of an
anti-correlation between $\alpha$ and $M(r<1.5h^{-1}\hbox{Mpc})$
(the HDM model is an extreme example of this).
Also shown as a big star in each of the panels of Figure (2)
is the data point for the Coma cluster of galaxies,
where the Coma cluster mass within the Abell radius,
$M(r<1.5h^{-1}\hbox{Mpc}) = 6.5\times 10^{14}h^{-1}\hbox{M}_\odot$,
is from the X-ray determination by Hughes (1989),
and the asymptotic slope of the mass distribution
in the Coma cluster ($\alpha=0.27$)
is adapted from The \& White (1986).
It is somewhat premature to make a definite conclusion based
only on one data point
of the Coma cluster,
but if one were forced to choose among models,
it seems that the two open models (panels c and d)
fare well in producing Coma-like clusters,
but the two flat models (panels a and b)
appear to produce much more extended clusters for masses like that
of the Coma.
A reasonable, physically plausible amount of bias
of galaxy distribution over mass on the relevant scales,
will not significantly alter these remarks.
A more realistic comparison requires to follow
the galaxy motion as well as the dark matter motion
in multi-component simulations, which at present are
prohibitively expensive.

To explore this further, we show, in another way,
the mass distribution in Figure (3),
where the abscissa is the
ratio of mass within a sphere
of radius $1.5h^{-1}$Mpc to that within a sphere of radius
$1.0h^{-1}$Mpc and the ordinate is the
ratio of mass within a sphere of radius $3.0h^{-1}$Mpc to that within a sphere
of radius $1.5h^{-1}$Mpc. Also shown as big solid dots are what one
should have,
if the density profile were isothermal.
We find that most of clusters
in all three CDM models tend to have mass distributions
more concentrated than isothermal distribution
(with some small fraction of clusters being exceptions).
But the HDM model shows just the opposite, again due to
the primary effect of fragmentation process in this model.

In summary, there are two factors which are important
to the mass distributions.
The first is the power ($P_k$) on relevant
scales ($\lambda \le 16h^{-1}$Mpc,
assuming that the power on larger scales is the same);
less power on the relevant scales tends
to point to the direction of an anti-correlation between $\alpha$
and
$M(r<1.5h^{-1}\hbox{Mpc})$,
i.e., to make
small mass clusters more extended
and massive clusters more concentrated.
The second factor is the mean density of the universe, $\Omega_0$;
less merging in a lower $\Omega_0$ universe tends to make
massive clusters more concentrated and less massive ones
more extended.
CDM-like (hierachical) models produce
density distributions
of most of the clusters
more concentrated than the isothermal distribution;
on the contrary, HDM-like (pancaking) models
produce
density distributions
of most of the clusters
more extended than the isothermal distribution.
The dependence on $P_k$ and $\Omega_0$
of the correlation between
$\alpha$ and $M(r<1.5h^{-1}\hbox{Mpc})$ as well the absolute
amplitude of $\alpha$
might provide
a way to decipher the initial
power on the relevant scales ($\lambda\le 16h^{-1}$Mpc)
and/or $\Omega_0$ of our universe,
when more data on the cluster mass distributions
becomes available.

\medskip
\centerline{\it 3.2 Cluster Internal Velocity Distributions}
\medskip
We now turn to the velocity distributions.
Figures (4,5,6,7) show the averages, $\langle \eta_m(r)\rangle$,
of the four normalized
velocity dispersions (as a function of cluster-centric distance)
defined as
$$\eqalignno{\eta_{m}(r) & \equiv \sigma_{m} (r)/\langle \sigma_{m}\rangle
\quad, \quad &(\the\eqtno )\cr}$$
\advance\eqtno by 1
where $\sigma_m$
is the velocity dispersion and
$m=(1d,r,t,||)$
for (1d, radial, tangential, line-of-sight), respectively.
The one-dimensional velocity dispersion is defined as
$$\eqalignno{\sigma_{1d} &\equiv {1\over \sqrt{3}} \sqrt{\sigma_x^2 +
\sigma_y^2+\sigma_z^2}\quad. \quad &(\the\eqtno )\cr}$$
\advance\eqtno by 1
The radial velocity dispsersion, $\sigma_r$,
in a shell $r\rightarrow r+\Delta r$,
is relative to the shell,
i.e., the infall or outflow velocity of the shell is removed before
calculating the velocity dispersion.
The line-of-sight velocity dispersion in each
cylindrical shell ($0.2h^{-1}$Mpc thick,
which cuts through the sphere of radius $3h^{-1}$Mpc)
is computed by averaging over all the lines of sight in
the shell.
We define
$$\eqalignno{\langle \sigma_{m}\rangle &\equiv {1\over R}\int_0^{R}
\sigma_{m}(r)dr \quad, \quad &(\the\eqtno )\cr}$$
\advance\eqtno by 1
where $R=3h^{-1}$Mpc.
The average, $\langle\eta_m(r)\rangle$,
is over
all the clusters with masses greater than $10^{13}h^{-1}M_\odot$
(solid line: number-weighted,
dotted line: mass-weighted) in each model.

In Figure (4) we see that the three CDM models
[panels (a,c,d)]
have a decreasing 1-d velocity
dispersion on scales from $1$ to $3h^{-1}$Mpc.
On the other hand, the HDM model [panel (b)]
shows an increasing 1-d velocity
dispersion with scale,
which is consistent with its
mass distribution (see \S 3.1).
Figures (5,6) show the normalized
radial and tangential velocity dispersions
(solid line: number-weighted, dotted line: mass-weighted).
We see similar results as in Figure (4).

Figure (7) shows the line-of-sight velocity dispersion as a function
of projected distance.
We see that the number-weighted
line-of-sight velocity dispersions (solid lines)
have weak, monotonically increasing distributions
(in terms of projected cluster-centric distance).
But the mass-weighted ones tend to tilt
clockwise (i.e., to the direction of becoming decreasing functions
in terms of scale).
Let us take the observed line-of-sight velocity
information on the Coma cluster of galaxies
to make a comparison with these four models.
Figure (8) shows the line-of-sight velocity dispersion
averaged over the clusters with masses in the range
from $3\times 10^{14}h^{-1}\hbox{M}_\odot$ to
$1\times 10^{15}h^{-1}\hbox{M}_\odot$.
Also shown are the observed data points (as solid dots)
for the Coma cluster compiled from Table (2) of Kent and Gunn (1982).
We see that the $\Omega_0=1$ CDM model (panel a)
and $\Omega_0=0.2$ CDM model (panel c)
provide good fits to observations
from $r=0.5h^{-1}$Mpc (below which
our simulation is not reliable) to $3.0h^{-1}$Mpc.
The reason that the panel (d) shows a lower
velocity dispersion than that in panel (c)
is that there are no clusters having mass greater than
$4\times 10^{14}h^{-1}\hbox{M}_\odot$ in the simulated box of model 4
due to limited simulation boxsize.
More quantitative comparison is not possible
given both the uncertainty of the mass of the Coma cluster
and the uncertainty of velocity bias of galaxies over matter.
But it appears that the CDM-like models are viable.
In contrast, the HDM model (panel b) provodes
a poor fit noting that the shape of the computed velocity dispersion
is very different from the observed counterpart.

Next, we study the issue
of anisotropy of the velocity distributions.
Let us first define two measures for the anisotropy of the velocity
distribution:
$$\eqalignno{\epsilon_{||} (r)&\equiv
\sqrt{\sigma_{||,y}^2(r)+\sigma_{||,z}^2(r)\over 2\sigma_{||,x}^2(r)}
\quad \quad &(\the\eqtno )\cr}$$
\advance\eqtno by 1
and
$$\eqalignno{\epsilon_{rt} (r)&\equiv {\sigma_r(r)\over\sigma_t(r)}
\quad. \quad &(\the\eqtno )\cr}$$
\advance\eqtno by 1
The former measures the anisotropy
among the three
orthogonal
Cartesian
directions ($x,y,z$),
and the latter measures anisotropy between the radial and tangential
velocity dispersions (by a local observer).
Figures (9,10) show
$\epsilon_{||} (r)$
and
$\epsilon_{rt} (r)$, respectively,
as a function of projected cluster-centric distance, $r_p$,
and 3-d cluster-centric distance, $r$, respectively,
averaged over
all the clusters with masses greater than $10^{13}h^{-1}$M$_\odot$
(equally weighted).
We see, in Figure (9),
that all the models have a mean of $\epsilon_{||}$
around unity, i.e., isotropic,
but they show quite significant amount
of scatters around the mean, 40-50\%.
Examination of Figure (10) (solid lines: number-weighted,
dotted lines: mass-weighted) reveals
that the velocity dispersions in all models
are anisotropic between radial and tangential components.
It shows that
the tangential velocity dispersions are larger
than the corresponding radial velocity dispersion
by a factor of 1.2-1.6 in the CDM models
and 1.3-2.0 in the HDM model.
In all models the ratio of radial to tangential velocity
dispersions show a decrease from $0.5h^{-1}$Mpc
to $3.0h^{-1}$Mpc for massive clusters
(smaller clusters tend to show a minimum for that ratio
around $1.5-2.0h^{-1}$Mpc in the CDM models).

Let us now turn to the infall issue.
Figure (11) shows the radial velocity of each cluster-centric shell
relative to an observer at the cluster center as a function of
cluster-centric distance,
averaged over all the clusters with
$M(r<1.5h^{-1}\hbox{Mpc})>10^{13}h^{-1}M_\odot$
(solid line: number-weighted,
dotted line: mass-weighted).
Note that in a homogeneous uniform universe $v_r=Hr$.
We see, as expected, that flat ($\Omega_0=1$)
models
have larger turnaround radii ($R_{ta}$), where $v_r=0$,
than lower $\Omega_0$ models,
and massive clusters have larger turnaround radii
than poorer clusters.
Figure (12) shows
$R_{ta}$ as a function
of the mass inside the turnaround radius, $M_{ta}$ (solid dots).
Also shown as the solid lines
are the least-square fitting curves with
the following formula:
$$\eqalignno{ \log_{10} R_{ta} &= a + b \log_{10} M_{ta}
\quad, \quad &(\the\eqtno )\cr}$$
\advance\eqtno by 1
where $R_{ta}$ is in $h^{-1}$Mpc and $M_{ta}$ is in $h^{-1}\hbox{M}_\odot$.
The values of $a$ and $b$ of the four
fits are tabulated as row 13 in Table 1.
It is interesting that in all the four models
the values of $a$'s and $b$'s are close
and can be represented by
$$\eqalignno{ a &= -5.2\pm 0.2 ~~~~\hbox{and}~~~~ b=0.40\pm 0.02
\quad. \quad &(\the\eqtno )\cr}$$
\advance\eqtno by 1
It is very instructive to look closely
at the three CDM models.
We see that panels (c) and (d)
look very similar
[note that the only difference between these two models
is the power $P_k$: panel (d) has more power than panel (c)].
This means that plausible variations of $P_k$ have negligible
effect on the turnaround radius for a given cluster mass.
However, we notice that the points in
panel (a) are always below those in panel (c) [and (d)]
for a fixed mass,
which seems surprising at first.
The reason is the following.
In order to accumulate the initially uniformly distributed matter
into islands of matter,
larger spatial volumes of matter
are required in a lower $\Omega_0$ model
than in a higher $\Omega_0$ model to reach
the same masses.
The countervailing factor is that a lower $\Omega_0$
tends to brake the building-up process,
thus reduces the infall.
But in the models we examine here, $\Omega_0=1$ versus $\Omega_0=0.2$ models,
it seems that the former factor is primary.
Note that, if the former factor were the only one,
the ratio of the two turnaround radii
of panel (a) to panel (c)
at a fixed mass should be $(0.2/1.0)^{1/3}=0.58$ whereas
we find the actual ratio is $0.79$.
We would like to stress that,
at a fixed cluster-centric distance or at a fixed
overdensity,
high $\Omega_0$ models have larger
turnaround radii than low $\Omega_0$ models (for a similar normalization,
e.g., $\sigma_8$), as was shown
in Figure (11); this topic has been discussed in Cen (1994)
focusing on the kinematic behaviors of the Local Supercluster.
Another related topic has been discussed
by Peebles \etal (1989) with regard to
the Local Group dynamics, concluding that a flat model
is consistent with observed motions in the local group.
We note, however,
that the $\Omega_0=1$ model which fits the observed infall motion
in the Local Group (our Galaxy relative to the Andromeda Nebula)
requires $H=80$km/s/Mpc, as found by Peebles \etal (1989),
giving the age of the
universe of 8.2 billion years, which seems too short.
If one presses hard on the age issue,
a lower $\Omega_0$ model might provide a better fit,
since a lower $\Omega_0$ model
will have a smaller infall motion at a fixed separation
(the Galaxy relative to the Andromeda),
which seems to point
to the right direction of having a longer age.
Very high resolution simulations, taking into account
of large scale environmental effects,
are needed
before we can make more quantitative assessments
of what ranges of $\Omega_0$ and $H_0$ fit.

Finally, we look at the relation between
mass overdensity inside turnaround radius, $\delta_{ta}$,
and $M_{ta}$.
Figure (13) shows $\delta_{ta}$ as a function of $M_{ta}$.
We see that the flat models have a lower overdensity than the open
models ($\sim 4-10$ vs. $\sim 10-30$) within the turnaround
radius.
Also shown as the horizonal, dashed lines (in panels a and b)
are the analytic prediction for an isolated spherical collapse
case in an Einstein-de Sitter universe (\cf Peebles 1980).
It is interesting that this analytic calculation gives
a reasonable fit to the fully nonlinear, non-spherical simulations
(panels a and b).
The three CDM models all show
a trend that massive clumps are likely to have lower values
of overdensity inside the turnaround spheres than small mass clumps.
We fit the open circles in panels (a,c,d)
by the following formula:
$$\eqalignno{\log_{10}\delta_{ta} &= c + d \log_{10} M_{ta}\quad, \quad
&(\the\eqtno )\cr}$$
\advance\eqtno by 1
where $M_{ta}$ is in $h^{-1}\hbox{M}_\odot$.
The values of c and d are listed as row 14
in Table 1 (we do not fit for the HDM model).

\medskip
\centerline{\it 3.3 The Isothermal Model}
\medskip
It is frequently assumed that the cluster
density distribution can be approximated by
an isothermal profile (Peebles 1993).
This leads to a way to estimate the cluster
mass given its velocity information as follows.
$$\eqalignno{M_{ISO}(<r)&= {2v_{1d}^2(<r) r\over G}
\quad, \quad &(\the\eqtno )\cr}$$
\advance\eqtno by 1
where $v_{1d}(<r)$ is the 1-d velocity dispersion
within a sphere of radius $r$;
$G$ is the gravitational constant.
We now examine this issue in numerical simulations
assuming that observed velocity information is error-free.
Figures (14,15,16) show the ratios of
the derived masses by assuming an isothermal distribution [equation (8)]
to the true masses with spheres of radii
$1.0h^{-1}$Mpc,
$1.5h^{-1}$Mpc
and
$3.0h^{-1}$Mpc, respectively.
We see that, in all three scales for all the models,
there is a rather significant scatter of estimated masses.
More seriously,
with a radius $r=1h^{-1}$Mpc, the isothermal approximation
seems to underestimate the mean by 20-30\% in the CDM models,
although the scatter is still large so in some case they agree within scatter.
On the scale $r=1.5h^{-1}$Mpc, the isothermal approximation
still underestimates the true masses
by 10-20\% in the CDM models but the agreement is better
especially for poorer clusters ($M<10^{14}h^{-1}M_\odot$).
Then on the scale $r=3.0h^{-1}$Mpc, the isothermal approximation
overestimates the true masses
by 10-30\% in the CDM models.
The HDM model shows that
the isothermal model typically
overestimates
the masses for scales $r<1.5h^{-1}$Mpc but
underestimates
the masses for scales $r>1.5h^{-1}$Mpc.
It seems that there is a
scale around $2h^{-1}$Mpc where, accidentally,
one may be able to get the right mass (on average)
using the isothermal approximation.

Combining this information with the anisotropy
found in Figure (9) ($\epsilon_{||}$),
we conclude that
the isothermal approximation tends to underestimate the mean
of the true masses within the Abell radius by
10-30\% with a scatter of $\sim 50\%$ around the
estimated mean (in the three hierachical models).

\bigskip
\medskip
\centerline{4. CONCLUSIONS}
\bigskip
\medskip
Our main conclusions can be summarized
as four points
with regard to the mass distribution, velocity
distribution, infall motion, and the isothermal model.

(1) We find,
by isolating the effects of $\Omega_0$
and $P_k$ (on the relevant scales, $\lambda\le 16h^{-1}$Mpc)
through a series of model simulations,
that both $\Omega_0$ and $P_k$
are important to the mass distributions in clusters of galaxies.
In the present study, we focus on the mass distributions
in the outskirts of clusters ($r=0.5-3h^{-1}$Mpc).
Our main conclusion on this issue is assembled through three
related points.
First, there is a correlation between
power, $P_k$ (on the relevant scales),
and density profiles of massive mass clusters;
more power tends to point to the direction
of a correlation between $\alpha$ and $M(r<1.5h^{-1}$Mpc)
[see equation (1) for definitions],
i.e., massive clusters being more extended and
small mass clusters being relatively concentrated.
Second,
a lower $\Omega_0$ universe tends to produce
relatively concentrated massive clusters and relatively
extended small mass clusters
compared
to their counterparts in a higher $\Omega_0$ model with the same power.
Third,
models with little (initial) small scale power, such as the HDM model,
tend to produce more extended mass distribution for
most of the clusters than the isothermal distribution.
But the CDM-like models show mass distributions of most
of the clusters more concentrated than the isothermal distribution.

X-ray observations, such as ROSAT and future satellite missions,
and observations of gravitational lensing of distant galaxies
by foreground clusters, producing effects such as coherent
ellipticity (Miralda-Escude 1991;
Blandford \etal 1991;
Kaiser 1992),
may provide useful information
on the mass distribution in and around clusters.
In fact, they are providing us with
some new, interesting observational results;
see Bonnet \etal (1994),
Mellier \etal (1994),
Fahlman \etal (1994),
Tyson (1994),
Dahle \etal (1994),
Smail \etal (1994)
for the lastest observational work on this subject.
Comparison between observations and detailed model computations
could yield some interesting constraints on
the (initial) power of the density fluctuations on scales
$\lambda \le 16h^{-1}$Mpc
(where linear extrapolation is invalid)
and/or $\Omega_0$.

(2) With regard to
the velocity distribution, we
divide our conclusion into two points.
First, in $0.5<r<3.0h^{-1}$Mpc region,
velocity dispersions of four
components, (1d,radial, tangential,$||$),
show decreasing distributions as a function
of cluster-centric distance in the three CDM models;
but the HDM models shows just the opposite: weakly
increasing velocity dispersions outwards.
The CDM models can reasonably fit the observed galaxy velocity
dispersions in the Coma cluster of galaxies but
the HDM provides a poor fit.
Second, while the velocity dispersions among the three Cartesian
directions are isotropic,
a large scatter (40\%) exists in all models.
We find that for the scales $0.5<r<3.0h^{-1}$Mpc,
the tangential velocity dispersion
is always larger than
the radial component
by a factor of 1.2-1.6 in the CDM models
and 1.3-2.0 in the HDM model.
In all models the ratio of radial to tangential velocity
dispersions show a decrease from $0.5h^{-1}$Mpc
to $3.0h^{-1}$Mpc for massive clusters
(smaller clusters tend to show a minimum for that ratio
around $1.5-2.0h^{-1}$Mpc in the CDM models).

(3) The relation between the turnaround radius and
the mass within that radius
can be approximated by
$\log_{10}R_{ta} = a + b \log_{10} M_{ta}$
where $a=-5.2\pm 0.2$, $b=0.40\pm 0.02$,
$R_{ta}$ is in $h^{-1}$Mpc and $M_{ta}$ is
in $h^{-1}\hbox{M}_\odot$ (valid for all the models examined here).
Lower $\Omega_0$ models are found to have
larger turnaround radius for a fixed mass clump
than high $\Omega_0$ models;
this conclusion is insensitive to $P_k$.
The relation between the overdensity inside the turnaround
radius and the mass within that radius
is fitted by
$\log_{10}\delta_{ta} = c + d \log_{10} M_{ta}$
and values of $c$ and $d$ are listed as the last row in Table 1.
We show that the isolated spherical collapse model
in an Einstein-de Sitter universe,
having $\delta_{ta}=9\pi^2/16=5.55$,
gives a fair fit to results ($\sim 4-10$)
of the nonlinear, non-spherical simulations
performed here (see Figure 13).
Lower $\Omega_0$ models have considerably higher
$\delta_{ta}$, $\sim 10-30$.
All models show a trend that massive clumps have
lower values of overdensity than
small mass clumps.

(4) The isothermal approximation (\cf equation 10)
tends to underestimate
the true masses within the Abell radius by
10-30\% with a scatter of $\sim 50\%$ around the
estimated mean (in the three hierachical models).
Accidentally, it seems that
within a sphere of radius $\sim 2h^{-1}$Mpc,
one may be able to get the right mass (on average, in all
the models)
using the isothermal approximation but the scatter
around the mean still exists.

I would like to thank
especially
Jerry Ostriker for encouragement and many useful suggestions,
and Michael Strauss for a careful reading of
the manuscript and many instructive comments and suggestions.
Discussions with Rich Gott are also very helpful.
It is a pleasure to acknowledge the help of NCSA
for allowing me to use their Convex-240 supercomputer.
This research is supported in part by NASA grant
NAGW-2448, NSF grant AST91-08103 and HPCC, NSF grant ASC-9318185.

\vfill\eject

\centerline{REFERENCES}
\bigskip
\medskip
\refset
Bahcall, N.A. 1988, ARAA, 26, 631
\smallskip
\refset
Bahcall, N.A., \& Cen, R.Y., 1992a, ApJ(Letters), 398, L81
\smallskip
\refset
Bahcall, N.A., \& Cen, R.Y., 1993, ApJ(Letters), 407, L49
\smallskip
\refset
Blandford, R.D., Saust, A.B., Brainerd, T., \& Villumsen, J.V.
1991, MNRAS, 251, 600
\smallskip
\refset
Bonnet, H., Fort, B., Kneib, J.P., Mellier, Y., \& Soucail, G. 1994,
A\& A, 280, L7
\smallskip
\refset
Carlberg, R.G., Couchman, H.M.P., \& Thomas, P.A. 1990,
ApJ(Letters), 352, L29
\smallskip
\refset
Carlberg, R.G., \& Dubinski, J 1991, ApJ(Letters), 369, 13
\smallskip
\refset
Cen, R.Y. 1992, ApJS, 78, 341
\smallskip
\refset
Cen, R.Y., \& Ostriker, J.P. 1992, ApJ(Letters), 399, L113
\smallskip
\refset
Cen, R.Y., \& Ostriker, J.P. 1993a, ApJ, 417, 404
\smallskip
\refset
Cen, R.Y., \& Ostriker, J.P. 1993b, ApJ, 417, 415
\smallskip
\refset
Cen, R.Y. 1994, ApJ, 424, 0
\smallskip
\refset
Dahle, H., Maddox, S.J., \& Lilje, P.B. 1994, in preparation
\smallskip
\refset
Evrard, A.E, Summers, F.J., Davis, M. 1994, ApJ, 422,11
\smallskip
\refset
Fahlman, G.G., Kaiser, N., Squires, G., \& Woods, D. 1994, preprint
\smallskip
\refset
Henry, J.P., \& Arnaud, K.A. 1991, ApJ, 372, 400
\smallskip
\refset
Hockney, R.W., and Eastwood, J.W. 1981, ``Computer Simulations
Using Particles", McGraw-Hill, New York.
\smallskip
\refset
Hughes, J.P. 1989, ApJ, 337, 21
\smallskip
\refset
Kaiser, N. 1992, ApJ, 388, 272
\smallskip
\refset
Katz, N., Hernquist, L., \& Weinberg, D.H. 1992, ApJ(Letters), 399, L109
\smallskip
\refset
Kent, S.M., \& Gunn, J.E. 1982, AJ, 87, 945
\smallskip
\refset
Mellier, Y., Fort, B., Bonnet, H., \& Kneib, J.P. 1994,
to appear in ``Cosmological Aspects of X-ray Clusters of Galaxies"
NATO Advanced Study Institute, W. Seitter et al. eds
\smallskip
\refset
Merritt, D. 1986, ApJ, 313, 121
\smallskip
\refset
Miralda-Escude, J. 1991, ApJ, 380, 1
\smallskip
\refset
Ostriker, J.P. 1993, ARAA, 31, 689
\smallskip
\refset
Peebles, P.J.E. 1970, AJ, 75, 113
\smallskip
\refset
Peebles, P.J.E. 1980, ``The Large-Scale Structure
of the Universe" (Princeton: Princeton University Press)
\smallskip
\refset
Peebles, P.J.E. 1993, ``Principles of Physical Cosmology"
(Princeton: Princeton University Press)
\smallskip
\refset
Peebles, P.J.E., \& Silk, J. 1988, Nature, 335, 601
\smallskip
\refset
Peebles, P.J.E., Melott, A.L., Holmers, M.R., \& Jiang, L.R. 1989, ApJ,
345, 108
\smallskip
\refset
Rood, H.J., Page, T.L., Kintner, E.C., \&
King, I.R. 1972, ApJ, 175, 627
\smallskip
\refset
Schechter, P. 1976, ApJ, 203, 297
\smallskip
\refset
Silk, J., Wyse, R.G. 1993, Phys. Report, 213, 295
\smallskip
\refset
Smail, I., Ellis, R.S., Fitchet, M.J., \& Edge, A.C. 1994, in preparation
\smallskip
\refset
The, L.S., \& White, S.D.M. 1986, AJ, 92, 1248
\smallskip
\refset
The, L.S., \& White, S.D.M. 1988, AJ, 95, 15
\smallskip
\refset
Tyson, J. 1994, in proceeding of Les Houches summer school
\smallskip
\refset
White, S.D.M. 1976, MNRAS, 177, 717
\smallskip
\refset
White, S.D.M., Efstathiou, G., \& Frenk, C.S. 1993, MNRAS,
\smallskip
\refset
\vfill\eject

\centerline{\ \ \ \ \ \ FIGURE CAPTIONS}
\bigskip
\medskip
\hsize=5.25truein
\hoffset=2.45truecm
\item{Fig. 1--}
The density distribution
as a function of cluster-centric distance
for a few typical (randomly chosen)
clusters in the four models
[panel (a) for model 1,
panel (b) for model 2,
panel (c) for model 3,
panel (d) for model 4, see Table 1;
this order will be maintained in the subsequent figures].
The dashed lines indicate the case for an isothermal sphere.

\item{Fig. 2--}
The relationship between $\alpha$ and $M(r<1.5h^{-1}$Mpc$)$
[\cf equation (1) for definitions]
is shown as open circles.
The solid line in each panel
is the linear least-square fit for the open circles
weighted by the inverse of the uncertainty of each power law fit
($\Delta\alpha$).
Also shown as a big star in each of the panels
is the data point for the Coma cluster of galaxies,
where the Coma cluster mass within the Abell radius
($M(r<1.5h^{-1}Mpc) = 6.5\times 10^{14}h^{-1}M_\odot$)
is from X-ray determination by Hughes (1989)
and the asymptotic slope of the mass distribution
in the Coma cluster ($\alpha=0.27$) from The \& White (1986).

\item{Fig. 3--}
The abscissa is the
ratio of mass within a sphere
of radius
$1.5h^{-1}$Mpc to that within a sphere
of radius
$1.0h^{-1}$Mpc and the ordinate is the
ratio of mass within a sphere
of radius
$3.0h^{-1}$Mpc to that within a sphere
of radius
$1.5h^{-1}$Mpc. Also shown as big solid dots are what one
should have if the density profile is isothermal.

\item{Fig. 4--}
The average of the normalized 1-d
velocity dispersions as a function of cluster-centric distance
[see equations (2,3,4) for definitions, \S 3.2].

\item{Fig. 5--}
The average of the normalized radial
velocity dispersions as a function of cluster-centric distance
[see equations (2,4) for definitions, \S 3.2].

\item{Fig. 6--}
The average of the normalized tangential
velocity dispersions as a function of cluster-centric distance
[see equations (2,4) for definitions, \S 3.2].

\item{Fig. 7--}
The average of the normalized line-of-sight
velocity dispersions as a function of projected cluster-centric distance
[see equations (2,4) for definitions, \S 3.2].

\item{Fig. 8--}
The line-of-sight
velocity dispersions as a function of projected cluster-centric distance
for the four models,
averaged over clusters with masses
in the range
from $3\times 10^{14}h^{-1}\hbox{M}_\odot$ to
$1\times 10^{15}h^{-1}\hbox{M}_\odot$.
Also shown as solid dots are the data points for
the Coma cluster taken from Table 2 of Kent \& Gunn (1982).

\item{Fig. 9--}
The velocity anisotropy measure,
$\epsilon_{||}$ [see equation (5) for definition],
among the Cartesian orthogonal directions,
as a function of cluster-centric distance, $r$,
averaged over
all the clusters with masses greater than
$10^{13}h^{-1}$M$_\odot$.

\item{Fig. 10--}
The radial and tangential velocity anisotropy measure,
$\epsilon_{rt}$ [see equation (6) for definition],
as a function of cluster-centric distance,
averaged over
all the clusters with mass greater than $10^{13}h^{-1}$M$_\odot$
(solid curves: number-weighted,
dotted curves: mass-weighted).

\item{Fig. 11--}
The radial velocity (as observed by an observer
sitting at the center of the cluster)
as a function of
cluster-centric distance,
averaged over
all the clusters with mass greater than $10^{13}h^{-1}$M$_\odot$
(solid curves: number-weighted,
dotted curves: mass-weighted).

\item{Fig. 12--}
The turnaround radius, $R_{ta}$, as a function of the mass within
that radius, $M_{ta}$ (solid dots).
Also shown as solid lines are the least-square fits
[\cf equations (7,8) and Table 1].

\item{Fig. 13--}
The overdensity inside
the turnaround radius, $\delta_{ta}$,
as a function of the mass within that radius, $M_{ta}$ (open circles).
The dashed, horizontal lines in panels (a,b)
are the result for a
nonlinear, isolated spherical collapse model in
an Einstein-de Sitter universe (5.55).
Also shown as solid lines are the least-square
[\cf equation (9) and Table 1].

\item{Fig. 14--}
The ratio of
the derived mass by assuming an isothermal distribution [equation (10)]
to the actual computed mass with spheres of radius
$1.0h^{-1}$Mpc as a function of the true mass.

\item{Fig. 15--}
The ratio of
the derived mass by assuming an isothermal distribution [equation (10)]
to the actual computed mass with spheres of radius
$1.5h^{-1}$Mpc as a function of the true mass.

\item{Fig. 16--}
The ratio of
the derived mass by assuming an isothermal distribution [equation (10)]
to the actual computed mass with spheres of radius
$3.0h^{-1}$Mpc as a function of the true mass.

\vfill\eject

\hoffset=0.0truecm
\hsize=6.50truein
\centerline {{\bf Table 1.} Summary of the computed models}
\medskip
\begintable
\hfill Row \hfill|
\hfill Run \hfill|
\hfill 1 \hfill|
\hfill 2 \hfill|
\hfill 3 \hfill|
\hfill 4 \hfill\cr
\hfill  2 \hfill|
\hfill  Model \hfill|
\hfill  CDM  \hfill|
\hfill  HDM  \hfill|
\hfill  CDM  \hfill|
\hfill  CDM  \hfill\cr
\hfill  3 \hfill|
\hfill  $\Omega_0$ \hfill|
\hfill  $1$  \hfill|
\hfill  $1$  \hfill|
\hfill  $0.2$  \hfill|
\hfill  $0.2$  \hfill\cr
\hfill  4 \hfill|
\hfill  TF \hfill|
\hfill  CDM $\Omega_0=1$  \hfill|
\hfill  HDM $\Omega_0=1$  \hfill|
\hfill  CDM $\Omega_0=0.2$  \hfill|
\hfill  CDM $\Omega_0=1.0$  \hfill\cr
\hfill  5 \hfill|
\hfill  $h$  \hfill|
\hfill  $0.5$  \hfill|
\hfill  $0.5$  \hfill|
\hfill  $0.5$  \hfill|
\hfill  $0.5$  \hfill\cr
\hfill  6 \hfill|
\hfill  $n$  \hfill|
\hfill  $1$  \hfill|
\hfill  $1$  \hfill|
\hfill  $1$  \hfill|
\hfill  $1$  \hfill\cr
\hfill  7 \hfill|
\hfill  $\sigma_8$  \hfill|
\hfill  $1.05$  \hfill|
\hfill  $1.05$  \hfill|
\hfill  $1.05$  \hfill|
\hfill  $1.05$  \hfill\cr
\hfill  8 \hfill|
\hfill  $L(h^{-1}$Mpc)  \hfill|
\hfill  $100$  \hfill|
\hfill  $100$  \hfill|
\hfill  $100$  \hfill|
\hfill  $100$  \hfill\cr
\hfill  9 \hfill|
\hfill  $\Delta l(h^{-1}$Mpc)  \hfill|
\hfill  $0.2$  \hfill|
\hfill  $0.2$  \hfill|
\hfill  $0.2$  \hfill|
\hfill  $0.2$  \hfill\cr
\hfill  10 \hfill|
\hfill  $m_p(h^{-1}M_\odot)$  \hfill|
\hfill  $1.8\times 10^{10}$  \hfill|
\hfill  $1.8\times 10^{10}$  \hfill|
\hfill  $3.5\times 10^{9}$  \hfill|
\hfill  $3.5\times 10^{9}$  \hfill\cr
\hfill  11 \hfill|
\hfill  $N_p^*$  \hfill|
\hfill  $5555$  \hfill|
\hfill  $5555$  \hfill|
\hfill  $27777$  \hfill|
\hfill  $27777$  \hfill\cr
\hfill  12 \hfill|
\hfill  $n_{model}/n_{obs}^{**}$  \hfill|
\hfill  $19.0\pm 1.4$/$1.5_{0.5}^{+0.75}$  \hfill|
\hfill  $5.0\pm 0.7$/$1.5_{0.5}^{+0.75}$  \hfill|
\hfill  $0.9\pm 0.3$/$1.5_{0.5}^{+0.75}$  \hfill|
\hfill  $1.9\pm 0.4$/$1.5_{0.5}^{+0.75}$  \hfill\cr
\hfill  13 \hfill|
\hfill  $a/b$  \hfill|
\hfill  $-5.3/0.40$ \hfill|
\hfill  $-5.0/0.38$ \hfill|
\hfill $-5.2/0.40$ \hfill|
\hfill $-5.2/0.41$ \hfill\cr
\hfill  14 \hfill|
\hfill  $c/d$  \hfill|
\hfill  $4.4/-0.26$ \hfill|
\hfill ---\hfill |
\hfill $4.2/-0.21$ \hfill|
\hfill $4.0/-0.20$ \hfill
\endtable
$^*$ number of particles contained in a cluster with mass
$1.0\times 10^{14}h^{-1}M_\odot$.

$^{**}$ cumulative number density of clusters with
masses (with Abell radius)
greater than $1.8\times 10^{14}h^{-1}M_\odot$ for
the model and observation. The one $\sigma$ error bars for models
are Poissonian and the observation is from Bahcall \& Cen (1993).

\end